\documentclass[11pt]{ltxguide}

\usepackage{draftcopy}
\usepackage{amsmath}
\usepackage{amssymb}
\usepackage[left=30mm, right=35mm, top=20mm, bottom=20mm]{geometry}
\usepackage[numbers,sort & compress]{natbib}
\usepackage{bbold}
\usepackage{graphicx}
\usepackage{color}
\usepackage{float}
\usepackage{subfigure}
\usepackage{amsfonts}
\usepackage{txfonts}
\usepackage{fontenc}
\usepackage{tone}
\title{Heun-type solutions for the Dirac particle on the curved background of Minkowski space-times}

\author{\copyright~1999, Springer Verlag Heidelberg\\
   All rights reserved.}
\author{S Rahmani, H Panahi\thanks{Corresponding author (E-mail:
		t-panahi@guilan.ac.ir)}, %\orcidID{0000-1111-2222-3333} \and
 and
A Najafizade\\
{\small Department of Physics, Faculty of Basic Science, University of  Guilan, Rasht
	41635-1914, Iran.}}
\date{14 January 2023}

\begin{document}

\maketitle
\begin{abstract}
In the present paper, we study the Dirac equation in the background of Minkowski space-time on a light cone. With the help of the coupling of the radial parts, the system of 4 equations is reduced to two different second-order differential equations, which coincide with a particle in present potentials. It turns out that the central equation reduces to Heun-type equations and provides us with energy spectra. Also, it is shown that the Schr\"odinger equation like with a combination of the external field is related to a second-order differential equation. In another way to solve the problem, results are valid for a spinless charged particle in the context of the magnetic field.

	\textbf{Keywords:}Heun function, Dirac equation, curved space-time, light cone\\
%	\textbf{PACS Nos.}: {02.3}
\end{abstract}
\section{Introduction}
In recent years numerous articles in various branches of physics have addressed the issue of Heun functions \cite{ref1,ref2,ref3,ref4,ref5,ref6,ref7,ref8,ref9,ref10}. However, it seems that these functions are still more interesting since they contain accessible topics for physical problems in the future.
In 2013, Hortaçsu presented an interesting classification of the applications of Heun functions in solving different physical problems and reminded the expanse of these functions \cite{ref11}.
\par
In addition to the various applications of Heun functions \cite{ref4,ref5,ref6, ref7, ref8,ref9,ref10,ref11,ref12} in different branches of theoretical physics, such as physics quantum, optics, gravity, etc., we can also mention the traces of these equations in field theory problems with curved space-time models \cite{ref13,ref14,ref15,ref16,ref17}. 
 In this regard, Vieira and his colleagues have studied the Klein-Gordon equation in curved space with a  Kerr–Newman background and analytically have found exact solutions in terms of the confluent Heun functions for the radial and angular parts in \cite{ref18}. 
 The problem of a particle with spin 1/2 in the curved background space-time in the presence of the Coulomb potential, which reduces to a system of two first-order equations, has been derived in the conﬂuent Heun functions in Refs. \cite{ref19,ref20}.
  
In Ref. \cite{ref21}, the solutions of the Dirac equation have been studied the curved Nutku-Helicoid background in 4-dimensional space-times, which are in the trivial generalization to 5 dimensions the same as the double conﬂuent Heun-type. 
In another problem of curved space-times with a Minkowski background, a particle with spin 1 in the presence of an external Coulomb field has been investigated \cite{ref22} by rewriting a non-diagonal representation of tetrads.
\par
The aim of the present paper is to study the Dirac particle in the curved background of the Minkowski metric on the light cone. In order to obtain the eigenfunctions of this problem we use coupling method to reduce relations. The formalism expressed for solving the Dirac equation is based on a special trick to find the solutions of the Heun-type functions.
\par
The organization of the paper is as follows:
in the next section, we present the solutions of the Dirac equation equation in the background of the Minkowski metric on the light cone. Section \ref{sec2} is devoted to solutions of the Heun-type to the Dirac equation for two special potentials.
 In section \ref{sec3}, we discuss the solutions of the Dirac  to two forms in this background of spinless and show how to recover the expression of the confluent Heun function. The last section is dedicated to conclusions and avenues for further work. 
\section{Dirac equation in the constrained metric on the cone}\label{sec2}
Let us start with a metric in the spherically symmetric space-times. In simple transformation, evaluation for the line element taken in the 2-surface $S$ \cite{ref23,ref24}, reads by
\begin{equation}
	\mathrm{d}s^2=-2	\mathrm{d}u	\mathrm{d}v+\frac{1}{2}(u-v)^2(	\mathrm{d}\theta^2+\sin^2 \theta	\mathrm{d}\varphi^2),
\end{equation}
where $(\theta, \varphi)$ are the spherical coordinates with Riemannian metric $
\mathrm{d}\Omega^2=r^2(	\mathrm{d}\theta^2+\sin^2 \theta	\mathrm{d}\varphi^2)$ which is the intersection of two hypersurfaces in which constants of $u$ or $v$ are null cones (light-like). We have $-\infty <u\leq v< \infty$ for the retarded and advanced times, respectively, in which the radius at any value of $u$ and $v$ is given by  $r=(v-u)/\sqrt{2}\geq 0$. Choosing a non-diagonal tetrad according to
\begin{align}
	\begin{split}
	&e_{(0)}^\alpha =\frac{1}{\sqrt{2}}(1,0,0,1), \qquad e_{(1)}^\alpha =\frac{1}{\sqrt{2}}(1,0,0,-1),\\
	&e_{(2)}^\alpha =(0,\frac{\sqrt{2}}{u-v},0,0), \qquad e_{(3)}^\alpha =(0,0,\frac{\sqrt{2}}{u-v}\sin\theta,0),	
	\end{split}
\end{align}
therefore, Christoffel symbols are
\begin{align}
	\begin{split}
	&\Gamma^u_{\theta,\theta}=\frac{1}{2}(v-u), \qquad \Gamma^u_{\varphi,\varphi}=\frac{1}{2}(v-u)\sin^2(\theta),\\
	&\Gamma^v_{\theta,\theta}=\frac{1}{2}(u-v), \qquad \Gamma^v_{\varphi,\varphi}=\frac{1}{2}(u-v)\sin^2(\theta),\\
&\Gamma^\theta_{\varphi,\varphi}=-\cos\theta\sin\theta, \qquad \Gamma^\varphi_{\theta,\varphi}=\cot\theta,\\
	&\Gamma^\theta_{u,\theta}=\frac{1}{u-v}, \qquad \Gamma^\theta_{v,\theta}=\frac{1}{v-u},\\
		&\Gamma^\varphi_{u,\varphi}=\frac{1}{u-v}, \qquad \Gamma^\varphi_{v,\varphi}=\frac{1}{v-u}.\\
			\end{split}
\end{align}
The Dirac equation is written as \cite{ref24}
\begin{equation}\label{eq1}
[	i\hbar e^\mu_a\gamma^a(\partial_\mu+\Omega_\mu)-mc-V]\psi(x)=0, \qquad \mu=(u,v,\theta,\varphi)
\end{equation}
with
\begin{equation*}
\Omega_\mu=\frac{1}{8}\omega_{ab\mu}[\gamma^a,\gamma^b], \qquad \text{so that} \qquad \omega^a_{b\mu}=e^a_\nu\partial_\mu e^\nu_b +e^a_\nu e^\sigma_b \Gamma^\nu_{\sigma\mu}
\end{equation*}
where $\gamma^\mu$ denotes the Dirac matrices and $m$ represents the mass of the particle. The potential is considered in the form
\begin{equation}
	V=\gamma^\mu A_\mu+V_s I,
\end{equation}
where $A_\mu$ and $V_s$ are the vectors and scalar potentials, respectively. Units are chosen so that $c=\hbar=1$. The Dirac matrices satisfy
\begin{equation}
	\mathcal{J}^{ab}=[\gamma^a,\gamma^b].
\end{equation}
Therefore, eq. \eqref{eq1} takes the form
\begin{align}\label{eq2}
	\begin{split}
&\Big[\frac{i(u-v)}{2}\big[(\gamma^0+\gamma^3)\partial_u+(\gamma^0-\gamma^3)\partial_v\big]-
\frac{(u-v)}{2}\big[(\gamma^0+\gamma^3)A_u+(\gamma^0-\gamma^3)A_v\big]-
\big[\gamma^1 A_\theta+\frac{\gamma^2}{\sin \theta}A_\varphi\big]\\
&+i\big[\gamma^2\mathcal{J}^{32}-\gamma^1\mathcal{J}^{31}\big]
+i\gamma^1\partial_\theta+i\gamma^2\frac{\partial_\varphi-\mathcal{J}^{12}\cos\theta}{\sin\theta}-\frac{(m+V_sI)(u-v)}{\sqrt{2}}\Big]\psi=0.
	\end{split}
\end{align}
Our purpose is the solution of eq. \eqref{eq2} in the background of the Minkowski metric. The presence of the term $(v-u)$ throughout the reduced Dirac equation is a fundamental difficulty. Hence, let us consider the Dirac equation in coordinates $u$ and $v$  by so-called null coordinates as
\begin{equation}
u=\frac{1}{\sqrt{2}}(t-r),\qquad v=\frac{1}{\sqrt{2}}(t+r),
\end{equation}
label null reflects that $u$ and $v$ are constant along future and past null cones centered $r=0$. Subsequently, the infinitely for region $r\rightarrow \infty$ corresponds to taking one of the coordinates to infinity and another fixed. Substituting this transformation into eq. \eqref{eq2} leads to
\begin{align}\label{eq3}
	\begin{split}
	&\Big[-ir\gamma^0\partial_t+ir\gamma^3\partial_r+i\gamma^1\partial_\theta+i\gamma^2\frac{\partial_\varphi-\mathcal{J}^{12}\cos\theta}{\sin\theta}+\frac{1}{\sqrt{2}}\big[(\gamma^0+\gamma^3)A_u+(\gamma^0-\gamma^3)A_v\big]-\gamma^1 A_\theta\\
&	-\frac{\gamma^2}{\sin \theta}A_\varphi+(m+V_sI)r\Big]\psi=0.
		\end{split}
\end{align}
In the usual method, one can use the separation of variables method to obtain solutions at this stage. So, if we make the following ansatz for the spinor $\psi$ in terms of components of 4-vector in basis
\begin{equation}
	\psi(t,r,\theta,\varphi)=\begin{pmatrix}
		\Phi_1(t,r,\theta,\varphi)\\
		\Phi_2(t,r,\theta,\varphi)\\
		\Phi_3(t,r,\theta,\varphi)\\
		\Phi_4(t,r,\theta,\varphi)
	\end{pmatrix}=r^4e^{i\ell\varphi}e^{-	
	i\varepsilon t}\begin{pmatrix}
		\phi_1(r,\theta)\\
		\phi_2(r,\theta)\\
		\mathcal{X}_1(r,\theta)\\
		\mathcal{X}_2(r,\theta)
	\end{pmatrix},
\end{equation}
with $\varepsilon \in \mathbb{R}$ and $\ell \in \mathbb{Z}$ the energy and the azimuthal quantum number of the particle, respectively. The constraint equation \eqref{eq3} along the cone then leads to the following coupled linear system with first-order differential equations
\begin{align}
	\begin{split}
&\Big[-r\partial_r+\frac{i}{\sqrt{2}}(A_u-A_v)\Big]\phi_1+\Big[-\partial_\theta+\frac{\ell+A_\varphi}{\sin\theta}+2\cot\theta-iA_\theta\Big]\phi_2=i\Big[(\varepsilon+m+V_s)r-\frac{1}{\sqrt{2}}(A_u+A_v)\Big]\mathcal{X}_1,\\
&\Big[-r\partial_r+\frac{i}{\sqrt{2}}(A_u-A_v)\Big]\mathcal{X}_2+\Big[-\partial_\theta-\frac{\ell+A_\varphi}{\sin\theta}+2\cot\theta-iA_\theta\Big]\mathcal{X}_1=i\Big[(-\varepsilon+m+V_s)r+\frac{1}{\sqrt{2}}(A_u+A_v)\Big]\phi_2,\\
&\Big[r\partial_r-\frac{i}{\sqrt{2}}(A_u-A_v)\Big]\phi_2+\Big[-\partial_\theta-\frac{\ell+A_\varphi}{\sin\theta}+2\cot\theta-iA_\theta\Big]\phi_1=i\Big[(\varepsilon+m+V_s)r-\frac{1}{\sqrt{2}}(A_u+A_v)\Big]\mathcal{X}_2,\\
&\Big[r\partial_r-\frac{i}{\sqrt{2}}(A_u-A_v)\Big]\mathcal{X}_1+\Big[-\partial_\theta+\frac{\ell+A_\varphi}{\sin\theta}+2\cot\theta-iA_\theta\Big]\mathcal{X}_2=i\Big[(-\varepsilon+m+V_s)r+\frac{1}{\sqrt{2}}(A_u+A_v)\Big]\phi_1,
	\end{split}
\end{align}
which are now clearly integrable in the cone. Thus, with the assumption of taking the results in the product form of the radial and angular components 
\begin{equation}\label{eq4}
\Phi=\begin{bmatrix}
	\phi_1\\
	\phi_2\\
	\mathcal{X}_1\\
	\mathcal{X}_2
\end{bmatrix}=\begin{bmatrix}
S_1(r)T_1^+(\theta)\\
S_2(r)T_2^+(\theta)\\
S_3(r)T_1^-(\theta)\\
S_4(r)T_2^-(\theta)
\end{bmatrix}.
\end{equation}
Upon substitution of the separated wavefunction in \eqref{eq4}, one obtains 4 coupled radial equations $\{\phi_j,\mathcal{X}_j\}$ and the pair of angular components $T^+_j$ with $j=1,2$, respectively, that is
\begin{align}\label{eq7}
	\begin{split}
	&\Big[-r\partial_r+\frac{i}{\sqrt{2}}(A_u-A_v)\Big] S_1(r)-\lambda S_2(r)-i\Big[\varepsilon +m+V_s-\frac{1}{\sqrt{2}r}(A_u+A_v)\Big]rS_3(r)=0,\\
	&\Big[-r\partial_r+\frac{i}{\sqrt{2}}(A_u-A_v)\Big] S_4(r)-\lambda S_3(r)-i\Big[-\varepsilon +m+V_s+\frac{1}{\sqrt{2}r}(A_u+A_v)\Big]rS_2(r)=0,\\
	&\Big[r\partial_r-\frac{i}{\sqrt{2}}(A_u-A_v)\Big]S_2(r)+\lambda S_1(r)-i\Big[\varepsilon +m+V_s-\frac{1}{\sqrt{2}r}(A_u+A_v)\Big]rS_4(r)=0,\\
	&\Big[r\partial_r-\frac{i}{\sqrt{2}}(A_u-A_v)\Big]S_3(r)+\lambda S_4(r)-i\Big[-\varepsilon +m+V_s+\frac{1}{\sqrt{2}r}(A_u+A_v)\Big]rS_1(r)=0,
		\end{split}
\end{align}
and
\begin{align}\label{eq5}
	\begin{split}	
\Big(\partial_\theta+\frac{\ell+A_\varphi}{\sin\theta}-2\cot\theta+iA_\theta\Big)T_2^+=\lambda T_1^+,\\
	\Big(\partial_\theta-\frac{\ell+A_\varphi}{\sin\theta}-2\cot\theta+iA_\theta\Big)T_1^+=-\lambda T_2^+,
		\end{split}
\end{align}
where $\lambda$ is the separation constant. In which the angular parts $T^+_j$ related to the components of $T^-_j$ as: $T^+_j=T^-_j$ with $j=1,2$. Now assuming $A_\mu=0$, we start by examining the angular equation \eqref{eq5}, the equation of $T_j$  has the explicit form 
\begin{equation}\label{eq6}
\Big[\partial^2_\theta-4\cot\theta\partial_\theta+\frac{6-\ell(\ell\mp \cos\theta)}{\sin\theta}+\lambda^2-4\Big]=0.
\end{equation}
Using
$\cos\theta=2z-1, \ \sin^2\theta=4z(1-z)$,
the equation have the admissible solutions
\begin{equation}
	T_1(\theta)=N_n (\sin\theta)^{\ell+2}\cos\big(\tfrac{\theta}{2}\big)P_n^{(\ell+\tfrac{1}{2},\ell-\tfrac{1}{2})}(x),
\end{equation}
and
\begin{equation}
	T_2(\theta)=N'_n (\sin\theta)^{\ell+2}\sin\big(\tfrac{\theta}{2}\big)P_n^{(\ell-\tfrac{1}{2},\ell+\tfrac{1}{2})}(x),
\end{equation}
with
$x=-\cos\theta$
and where $P_n^{(\rho,\delta)}(x)$ denotes the Jacobi polynomials \cite{ref25}. These solutions derive the eigenvalue 
$\lambda=\left(n+\ell+\tfrac{1}{2}\right)$
with $n\in \mathbb{N}$.

We now examine the radial equations \eqref{eq7}. To simplify the calculations, we substitute  $\partial_rS_2$ and $\partial_rS_3$ second and fourth relations \eqref{eq7} to the first equation, and so on,  $\partial_rS_1$ and $\partial_rS_4$ first and third equations sit in the second equation. This expression in equation \eqref{eq7} gives us second-order, but uncoupled equations
\begin{align}
	\begin{split}
&\frac{\partial^2S_1}{\partial r^2}-\frac{\lambda}{r^2}S_2-\Big(\frac{\lambda^2}{r^2}+m^2+V_s^2+2mV_s-\varepsilon^2\Big)S_1=0,\\
&\frac{\partial^2S_2}{\partial r^2}-\frac{\lambda}{r^2}S_1-\Big(\frac{\lambda^2}{r^2}+m^2+V_s^2+2mV_s-\varepsilon^2\Big)S_2=0.
	\end{split}
\end{align}
Let us consider linear transformations on functions $S_1(r)$ and $S_2(r)$ as:
$S_1(r)+S_2(r)=\Gamma_1(r),\ S_1(r)-S_2(r)=\Gamma_2(r)$
depend on the radial variable
\begin{align}
	\begin{split}\label{eq8}
	&\Big[-\frac{\partial^2}{\partial r^2}+\frac{\lambda(\lambda+1)}{r^2}+(m^2+V_s^2+2mV_s-\varepsilon^2)\Big]\Gamma_1=0,\\
		&\Big[-\frac{\partial^2}{\partial r^2}+\frac{\lambda(\lambda-1)}{r^2}+(m^2+V_s^2+2mV_s-\varepsilon^2)\Big]\Gamma_2=0,
			\end{split}
\end{align}
then we see that only $\{\Gamma_1,\Gamma_2\}$ are coupled to each other, and also $\{\Gamma_3,\Gamma_4\}$.
Now, we solve one of these equations for the three models to determine the solutions for each arbitrary $n$.
\subsection{The Cornell potential}
\begin{figure}[h!]
	\begin{center}		\includegraphics[height=6cm,width=8cm]{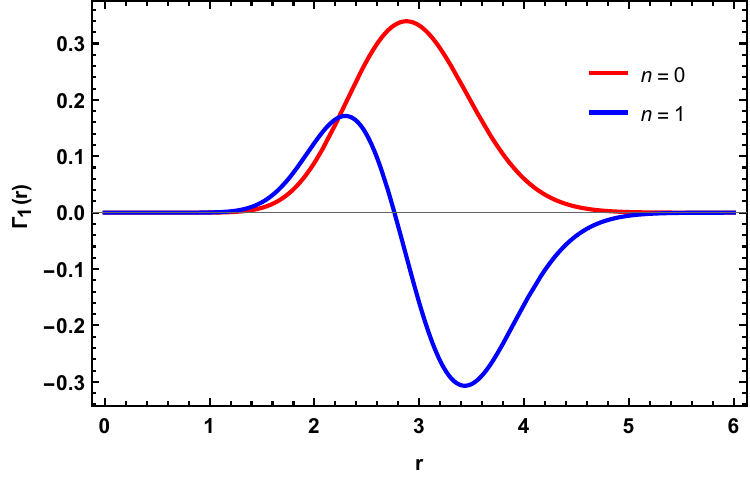}
	\end{center}
	\caption{Plots of the first few normalized wave functions $\Gamma_1(r)$ for the values $n=0,1$ corresponding to the solutions of the double-confluent Heun equation \eqref{eq12}.}\label{fig1}
\end{figure}

The Cornell potential is a combination of a Coulomb and linear potential. The Coulombic term describes a gluon exchange between the quark and its antiquark \cite{quark}. While, the linear term is related to the lattice QCD measurement \cite{qcd}. It is mainly used to describe the spectra of heavy quarkonium systems \cite{cornell1,cornell2} as
\begin{equation}\label{eq9}
	V_s(r)=-\frac{A}{r}+Br
\end{equation}
where $A$ and $B$ are constant coefficients. Substitution of \eqref{eq9} in eq. \eqref{eq8} yields
\begin{equation}\label{eq13}
	\frac{\mathrm{d}^2\Gamma_1}{\mathrm{d}r^2}+2(E-V_{eff}(r))\Gamma_1=0,
\end{equation}
which is the reduced one-dimensional stationary Schr\"odinger equation, such that the energy and effective potential read as
\begin{equation}\label{eq11}
	E=(\varepsilon^2-m^2)/2+AB, \qquad V_{eff}(r)=\frac{\lambda(\lambda+1)+A^2}{2r^2}-\frac{mA}{r}+\frac{B^2}{2}r^2+mBr
\end{equation}
Let us now consider an ansatz of the form
\begin{equation}\label{eq10}
	\Gamma_1(r)=\exp\Big[-\frac{B}{2}r^2-mr+\Lambda\log r\Big]f(r),
\end{equation}
By means of the function \eqref{eq10} into the Schr\"odinger equation \eqref{eq11}, we have
\begin{equation}\label{eq12}
	\frac{\mathrm{d}^2f(r)}{\mathrm{d}r^2}+\Big(\frac{\gamma}{r}+\delta+\epsilon r\Big)\frac{\mathrm{d}f(r)}{\mathrm{d}r}+\frac{\alpha r-q}{r}f(r)=0
\end{equation}
such that
\begin{align}
	\begin{split}
		&2\Lambda=1+\sqrt{1+4A^2+4\lambda^2+4\lambda}, \qquad \delta=-2m, \qquad \epsilon=-2B,\\
		&\alpha=\varepsilon^2+B(2A-2\Lambda-1), \qquad q=2m(A-\Lambda)
	\end{split}
\end{align}
Therefore, we obtained the energy eigenvalue expression for the Cornell potential as
\begin{equation}
	\varepsilon=\sqrt{B(2n+2\Lambda-2A+1)}
\end{equation}
Also, eq. \eqref{eq12} has the solution being expressed in terms of the bi-confluent Heun function \cite{my, bi-confluent, dong} as
\begin{equation}
	\Gamma_1(r)=N e^{\frac{\epsilon}{4}r^2+(\delta/2)r+\Lambda\log r}HeunB\Big[\gamma,\delta,\epsilon;\alpha,q,r\Big]	
\end{equation}
where  $N$ is a normalization constant. Fig. \ref{fig1} 
displays the bound-state wave functions of bi-confluent Heun type for some parameters.
%%%%%%%%%%%%%%%%%%%%%%%%%%%%%%%%%%%%%%%%%%
%%%%%%%%%%%%%%%%%%%%%%%%%%%%%%%%%%%
\subsection{The Kratzer potential}
\begin{figure}[h!]
	\begin{center}		\includegraphics[height=6cm,width=8cm]{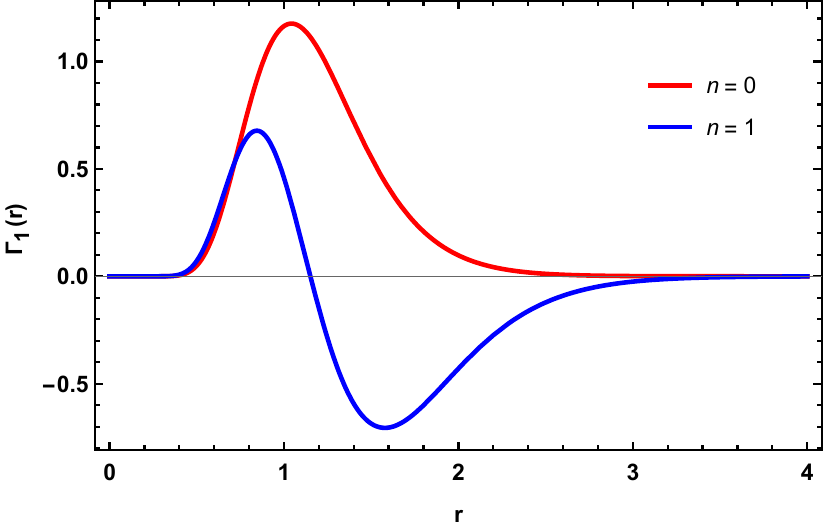}
	\end{center}
	\caption{Plots of the first few normalized wave functions $\Gamma_1(r)$ for the values $n=0,1$ corresponding to the solutions of the double-confluent Heun equation \eqref{eq15}.}\label{fig2}
\end{figure}

The Kratzer potential is one of the most important potentials for describing the vibrational and rotational structures of the diatomic molecules, and is discussed in three dimensions by Landau and Lifshitz \cite{ref26}. However, it can be the Coulomb plus inverse-square potential written as
\begin{equation}\label{eq16}
V_s(r)=\frac{C}{r^2}-\frac{D}{r},
\end{equation}
It should be noted that adding the inverse-square potential to the Coulomb potential can describe the electrons in the outer shell of an alkaline metal atom. After substitution of \eqref{eq8} in the original equation and in writing it to form \eqref{eq13}, we have 
\begin{equation}
E=(\varepsilon^2-m^2)/2, \qquad V_{eff}(r)=\frac{C^2}{2r^4}-\frac{CD}{r^3}+\frac{\lambda(\lambda+1)+2mC+D^2}{2r^2}-\frac{mD}{r}.
\end{equation}
Here, we again follow the another ansatz approach in the following form
\begin{equation}
\Gamma_1(r)=\exp\Big[\sqrt{m^2-\varepsilon^2}r+(1-D)\log r-\frac{C}{r}\Big]g(r).
\end{equation}
Hence, the Heun equation is given by,
\begin{equation}\label{eq15}
	\frac{\mathrm{d}^2g(r)}{\mathrm{d}r^2}+\Big(\frac{\gamma'}{r^2}+\frac{\delta'}{r}+\epsilon'\Big)\frac{\mathrm{d}g(r)}{\mathrm{d}r}+\frac{\alpha' r-q'}{r^2}g(r)=0,
\end{equation}
and corresponds to
\begin{align}
	\begin{split}
&\gamma'=2C, \qquad \delta'=2(1-D), \qquad \epsilon'=2\sqrt{m^2-\varepsilon^2},\\
&q'=\lambda(\lambda+1)-2C\Big(\sqrt{m^2-\varepsilon^2}-m\Big)+D, \qquad \alpha'=2(1-D)\sqrt{m^2-\varepsilon^2}+2Dm.
	\end{split}
\end{align}
For the derived eigenvalues
\begin{equation}
	\varepsilon=m\left[1-\frac{D^2}{(n+1-D)^2}\right]^{\frac{1}{2}}.
\end{equation}
The solution of the stationary Heun equation \eqref{eq15} for the presented potential \eqref{eq16} is thus explicitly written in terms of the double-confluent Heun function \cite{artur} as
\begin{equation}
\Gamma_1(r)=N' e^{\frac{\epsilon'}{2}r+\frac{\delta'}{2}\log r-\frac{\gamma'}{2r}}HeunD\Big[\gamma',\delta',\epsilon';\alpha',q;r\Big],	
\end{equation}
where  $N'$ is a normalization constant. Fig. \ref{fig2} 
displays the bound-state wave functions of double-confluent Heun type for some parameters.
%%%%%%%%%%%%%%%%%%%%%%%%%%%%%%%%%%%%%%%%%%
%%%%%%%%%%%%%%%%%%%%%%%%%%%%%%%%%%%
%%%%%%%%%%%%%%%%%%%%%%%%%%%%%%%%%
\section{Solutions of the Heun-type in an approach to the potential fields}\label{sec3}
In spherical coordinates, the Hamiltonian of a charged particle under the influence of an external magnetic field may be expressed as
\begin{align}
\Biggl\{&-i\gamma^0(\partial_t-ieA_t)+i\left(\gamma^3(\partial_r-ieA_r)-\frac{\gamma^1\mathcal{J}^{31}+\gamma^2\mathcal{J}^{32}}{r}\right)+\frac{1}{r}\left(i\gamma^1(\partial_\theta-ieA_\theta)+\gamma^2\frac{i(\partial_\varphi-ieA_\varphi)+\mathcal{J}^{21}\cos\theta}{\sin\theta}\right)\nonumber\\
&+mc\Biggr\}\psi=0,
\end{align}
where $A_\mu, \ \mu=t,r,\theta,\varphi$ is the 4-vector potential. After separating the variables and using the before section relations, we get four radial equations
\begin{align}
&(\partial_r-ieA_r)S_1+\frac{\lambda}{r}S_2+i(\varepsilon+m+\phi)S_3=0,\\
&(\partial_r-ieA_r)S_4+\frac{\lambda}{r}S_3-i(\varepsilon-m+\phi)S_2=0,\\
&(\partial_r-ieA_r)S_2+\frac{\lambda}{r}S_1-i(\varepsilon+m+\phi)S_4=0,\\
&(\partial_r-ieA_r)S_3+\frac{\lambda}{r}S_4+i(\varepsilon-m+\phi)S_1=0.
\end{align}
Here, for simplicity of calculations, we assume that the angular component of the potential is $A_\theta=A_\varphi=0$. Now let us consider the following condition for radial functions as
\begin{equation}
	S_1=S_2, \qquad S_4=-S_3,
\end{equation}
by taking $e A _t = \xi(r)$, we get
\begin{align}
	&\left[(\partial_r-ieA_r)+\frac{\lambda}{r}\right]S_1=-i(\varepsilon+m+\xi)S_3,\\
&\left[(\partial_r-ieA_r)-\frac{\lambda}{r}\right]S_3=-i(\varepsilon-m+\xi)S_1,
\end{align}
Substitution first function into the second function provides us with a second-order differential equation to the form
\begin{align}\label{eq20}
S''_1(r)-2ieA_r S'_1(r)&+\frac{1}{r\left(\varepsilon+\xi(r)+m\right)}	\left(rS'_1(r)-(ierA_r+\lambda)S_1(r)\right)\xi'(r)\nonumber\\
&+\left((\varepsilon+\xi(r))^2-m^2-e^2A_r^2-\frac{\lambda(\lambda-1)}{r^2}\right)S_1(r)=0.
\end{align}
Among the usual applications of the Dirac equation in an external field are those in which the spatial components $A_r$ vanish and $\xi(r)$ depends only on the magnitude $r$ of the position vector as happens in Ref. \cite{ref27}.  Therefore, equation \eqref{eq20} becomes simpler
\begin{align}
	S''_1(r)+\frac{1}{r\left(\varepsilon+\xi(r)+m\right)}	\left(rS'_1(r)-\lambda S_1(r)\right)\xi'(r)
	+\left((\varepsilon+\xi(r))^2-m^2-\frac{\lambda(\lambda-1)}{r^2}\right)S_1(r)=0.
\end{align}
Let us consider the problem of the Dirac equation in the presence of the external Coulomb field
$\xi(r)=-\frac{z}{r}$. With changing the variable
$x=\frac{m+\varepsilon}{z}r$, and using a substitution $S_1(x)=x^{-w}e^{-\tau x}h(x)$, 
in which
\begin{equation}
	 \qquad w=\frac{1}{2} (-1 + \sqrt{1 - 4 \lambda+ 4 \lambda^2 - 4 z^2-8\eta-4\eta \lambda}), \qquad \tau=\sqrt{z\eta(m-\varepsilon)}, \qquad \text{so that} \qquad \eta=\frac{z}{m+\varepsilon},
\end{equation}
so we get to the 
\begin{equation}
	\frac{\mathrm{d}^2h(x)}{\mathrm{d}x^2}+\Big(\frac{\gamma''}{x}+\frac{\delta''}{x-1}+\epsilon''\Big)\frac{\mathrm{d}h(x)}{\mathrm{d}x}+\frac{\alpha'' x-q''}{x(x-1)}h(x)=0,
\end{equation}
which is a confluent Heun differential equation and in within we have the parameters
\begin{align}
	\begin{split}
		&\gamma''=-2w, \qquad \delta''=-\eta, \qquad \epsilon''=2\tau,\\
		&\alpha''=-\eta(2\varepsilon z+\tau)-2\tau w, \qquad q''=-\eta(2\varepsilon z+w+\lambda)-2\tau w.
	\end{split}
\end{align}
Based on these parameter values, we find the confluent Heun functions $HeunC\Big[\gamma'',\delta'',\epsilon'',\alpha'',q'',x\Big]$ have the following forms \cite{dong1,artur1}:
\begin{equation}
	S_1(x)=N'' e^{\frac{\epsilon'}{2}r+\frac{\delta'}{2}\log r-\frac{\gamma'}{2r}}HeunC\Big[\gamma'',\delta'',\epsilon'';\alpha'',q'';x\Big],	
\end{equation}
where  $N''$ is a normalization constant.

One may also consider the situation when $\xi$ vanish and only the $A_r$ radial term remains, we have
\begin{align}\label{eq21}
	\left[\partial_r^2-2ieA_r \partial_r+\left(\varepsilon^2-m^2-e^2A_r^2-\frac{\lambda(\lambda-1)}{r^2}\right)\right]S_1(r)=0,
\end{align}
Writing now $\varepsilon=m+\kappa$, the $\varepsilon^2-m^2$ term of \eqref{eq21} becomes approximately $2m\kappa$ if $\kappa<<m$. Thus in the non-relativistic limit the energy $\kappa$ of the problem corresponds to a spinless charged particle in time-varying magnetic field presented in Ref. \cite{ref28}.
%%%%%%%%%%%%%%%%%%%%%%%%%%%%%%%%%
\section{Conclusions}
In this paper, we presented an approach to the method based on the usual tetrads for deriving the Dirac equation in the curved background on the light cone.
It turns out that  for null coordinates in Minkowski space-time, the equation \eqref{eq2} cannot coupled to have an analytic solution.
However, in changing null coordinates to spherical coordinates, the corresponding coupled equations are satisfied Heun-type functions depending on the chosen potential.
It is seen that the addition of a magnetic field acts in an almost similar way when they cause the differential equation to couple. It is shown that the confluent Heun equation may result in a suitable choice of scalar potential. Additionally, it is found that the known in the absence of a scalar potential,  may be recovered as a spinless charged particle in time-varying magnetic field.

It is important to note that some authors have used a  realization of tetrad concepts. However, their formalism is employed to solve the Dirac equation in various literature including: Cartan's formalism\cite{ref29}, the gauge
covariant formulation \cite{ref30}, the Newman-Penrose formalism \cite{ref31}, and the Kinnersley-like null frame \cite{ref32}, are different from those introduced.  
 Furthermore, quantum deformations have been shown \cite{ref33} to spaces of non-constant curvature can be considered.
\section*{Data availability statement}
All data that support the findings of this study are included within the article (and any supplementary files).

%{\color{red} \section*{ACKNOWLEDGMENTS}

\end{document}